\documentclass[useAMS,usenatbib,mnras]{mn2e}
\usepackage{aas_macros}
\usepackage[pdftex]{graphicx}
\usepackage{amsmath}
\usepackage{float}

\def\pasa{Publ. Astron. Soc. Australia}

\title[Young Pulsar Extreme Scattering Events]{Extreme Scattering Events Towards Two Young Pulsars}

\author[M. Kerr, W.~A. Coles, S. Johnston, C.~A. Ward, A.~V. Tuntsov, R. M. Shannon]
{M.~Kerr$^{1}$\thanks{E-mail: matthew.kerr@gmail.com},
  W.~A.~Coles$^{2}$,
  C.~A.~Ward$^{3,4}$,
  S.~Johnston$^{3}$,
  A.~V.~Tuntsov$^5$,
  \newauthor and R.~M.~Shannon$^{3}$ \\
$^{1}$Space Science Division, Naval Research Laboratory, Washington, DC
20375-5352, USA\\
$^{2}$ECE Department, University of California at San Diego, La
Jolla, CA 92093-0407, USA\\
$^{3}$CSIRO Astronomy and Space Science, Australia Telescope National
Facility, PO~Box~76, Epping NSW~1710, Australia\\
$^{4}$Sydney Institute for Astronomy, School of Physics, University of
Sydney, NSW 2006, Australia\\
$^{5}$Manly Astrophysics, 15/41-42 East Esplanade, Manly, NSW 2095,
Australia
}
\begin{document}

\date{Accepted 2017 November 22. Received 2017 November 22; in original form 2017 September 15}

\pagerange{\pageref{firstpage}--\pageref{lastpage}} \pubyear{2017}

\maketitle

\label{firstpage}

\begin{abstract}
We have measured the scintillation properties of 151 young, energetic
pulsars with the Parkes radio telescope and have identified two
extreme scattering events (ESEs).  Towards PSR~J1057$-$5226 we
discovered a three-year span of strengthened scattering during which
the variability in flux density and the scintillation bandwidth
decreased markedly.  The transverse size of the scattering region is
$\sim$23\,au, and strong flux density enhancement before and after the
ESE may arise from refractive focusing.  Long observations reveal
scintillation arcs characteristic of interference between rays
scattered at large angles, and the clearest arcs appear during the
ESE.  The arcs suggest scattering by a screen 100--200\,pc from the
earth, perhaps ionized filamentary structure associated with the
boundary of the local bubble(s).

Towards PSR~J1740$-$3015 we observed a ``double dip'' in the
measured flux density similar to ESEs observed towards compact
extragalactic radio sources.  The observed shape is consistent with
that produced by a many-au scale diverging plasma lens with electron
density $\sim$500\,cm$^{-3}$.  The
continuing ESE is at least 1500\,d long, making it the longest
detected event to date.

These detections, with materially different observational signatures,
indicate that well-calibrated pulsar monitoring is a keen tool for ESE
detection and ISM diagnostics.  They illustrate the strong r\^{o}le
au-scale non-Kolmogorov density fluctuations and the local ISM
structure play in such events and are key to understanding both their
intrinsic physics and their impact on other phenomena, particularly
fast radio bursts.
\end{abstract}


\section{Introduction}

The discovery of dramatic variations in the observed flux density of
the quasar 0954$+$658 \citep{Fiedler87}, dubbed an ``extreme
scattering event'', revealed an unexpected and poorly understood
property of the ionized interstellar medium (ISM).  If produced by
refraction \citep{Romani87}, the required column density gradients
imply densities $\geq$1000\,cm$^{-3}$ for symmetric distributions and
are consequently overpressured relative to the ambient ISM by three
orders of magnitude.  Various mechanisms, from local density
enhancements \citep{Walker98} to highly asymmetric structures like
current sheets \citep{Pen12} and circumstellar plasma streams
\citep{Walker17} have been proposed to alleviate this overpressure
problem, but the origin of the structure remains unclear.

A primary difficulty in advancing the field is the relative rarity of
such events.  They are typically discovered in archival data and thus
lack the contemporaneous VLBI/multi-frequency observations that help
probe the lensing structure.  Only recently was an ESE detected and
studied in real time via a monitoring campaign of
bright extragalactic AGN \citep{Bannister16}, and the wide frequency
coverage and VLBI imaging provided detailed insight into the electron
column density profile \citep{Tuntsov16}.

While dedicated campaigns like those of \citet{Bannister16} will
continue to turn up new ESEs, we point out pulsar monitoring as a
complementary approach.  Indeed, pulsar point sources are ideal for
detecting ESEs since the ESE signature is not convolved with a finite
source geometry.  Further, examination of differential delays in pulse
arrival times admits direct measurement of the electron column density
(dispersion measure, DM) towards the source.  \citet{Coles15}
identified two ESEs towards millisecond pulsars via their variations
in DM and scintillation strength, and there are likely a spectrum of
less extreme events in the dataset, making ISM studies a fruitful
offshoot of the effort to detect low-frequency gravitational waves by
pulsar timing arrays.  However, MSPs are much rarer than
unrecycled pulsars and so probe few lines of sight.  Thus, with an aim
towards discovering new ESEs, we searched over 1200 source-years of
data taken for the young pulsar timing program on the Parkes telescope
\citep[][``P574'']{Weltevrede10}.

We discovered two clear ESEs with dramatically different
observational signatures.  One, towards PSR~J1740$-$3015, is a smooth
modulation of the pulsar light curve reminiscent of the Fiedler et al.
event and suggestive of coherent lensing by a plasma structure.  The
other, towards PSR~J1057$-$5226, is more like the events reported by
\citet{Coles15}, detected through modification of diffractive
scintillation properties.  Although this latter type, observed
exclusively towards pulsars, have not been identified as ESEs in the
literature until recently, they too are
caused by au-scale inhomogeneities that cross the line of sight and
substantially alter/dominate the scattering properties.  It is
only in the addition of diffractive effects from the
pointlike sources that their observational signature differs, and
accordingly we classify both episodes as ESEs.

In \S\ref{sec:data}, we discuss the P574 observations and the
production of the dynamic spectra on which this analysis is based.  We
follow in \S\ref{sec:j1057} with a discussion of PSR~J1057$-$5226, its
scintillating nature, and the ESE detected towards it.  We do the same
for PSR~J1740$-$3015 in \S\ref{sec:j1740}.  In \S\ref{sec:discussion},
we offer interpretation of the observations.

\begin{table}
\caption{\label{tab:j1057_fiducial}Fiducial scattering model values for PSR~J1057$-$5226 for observing frequency 1369\,MHz and screen at $\zeta=1/2$.}
\begin{tabular}{lr}
\hline
Distance ($D$, pc)\dotfill & 400 \\ 
Dispersion Measure (DM, $e^{-1}$\,pc\,cm$^{-3}$) \dotfill & 30.1 \\ 
Transverse velocity (km/s) \dotfill & 80 \\
Fresnel radius (cm) \dotfill & \dotfill $6.5\times10^{10}$ \\
Decorrelation bandwidth ($B_d$, MHz) \dotfill & 50 \\
Decorrelation time-scale (s) \dotfill & 1000 \\
Diffractive scale ($s_d$, cm) \dotfill & $4.4\times10^9$ \\
Refractive scale (cm) \dotfill & $9.6\times10^{11}$ \\
Diffractive time-scale (s) \dotfill & $1100$ \\
Refractive time-scale (hr) \dotfill & $66$ \\
\end{tabular}
\end{table}

\begin{figure*}
  \includegraphics[angle=0,width=0.95\textwidth]{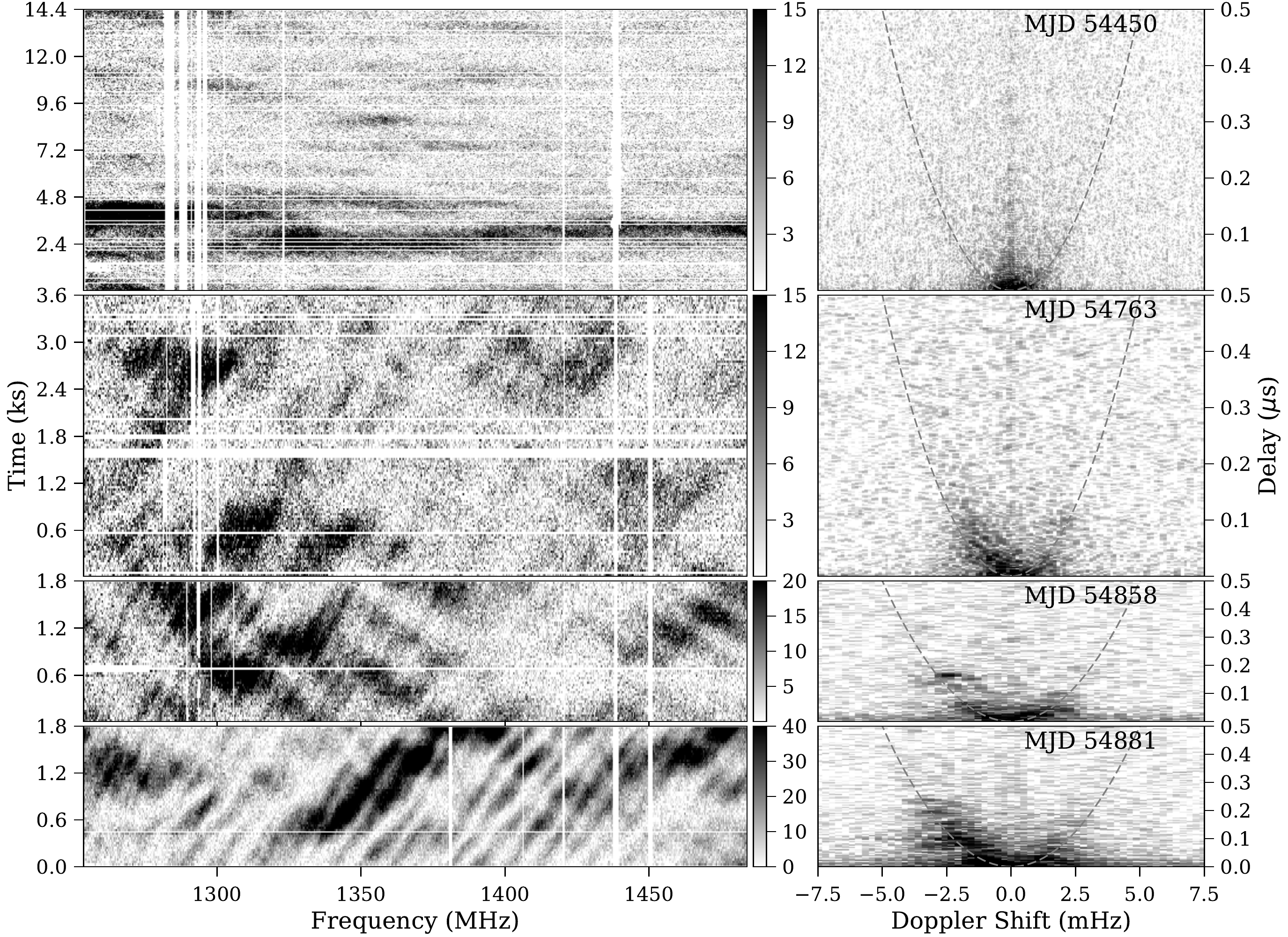}
  \caption{\label{fig:j1057_arcs_2}Dynamic and secondary spectra for
    observations of PSR~J1057$-$5226 in 2007 and 2008.  Left:
    Greyscale gives flux density in mJy.  Dynamic spectra generally
    have different durations and pixel scales.  Right: The secondary
    spectra are not to scale, but all share the same axes bounds.  The
    greyscale floor is set at the white noise level, and the maximum
three decades above.  The model arcs are drawn with $a=0.02$\,s$^3$,
the predicted value for our fiducial distance/velocity model (eqs.
\ref{eq:parab1} and \ref{eq:parab2})} \end{figure*}

\begin{figure*}
  \includegraphics[angle=0,width=0.95\textwidth]{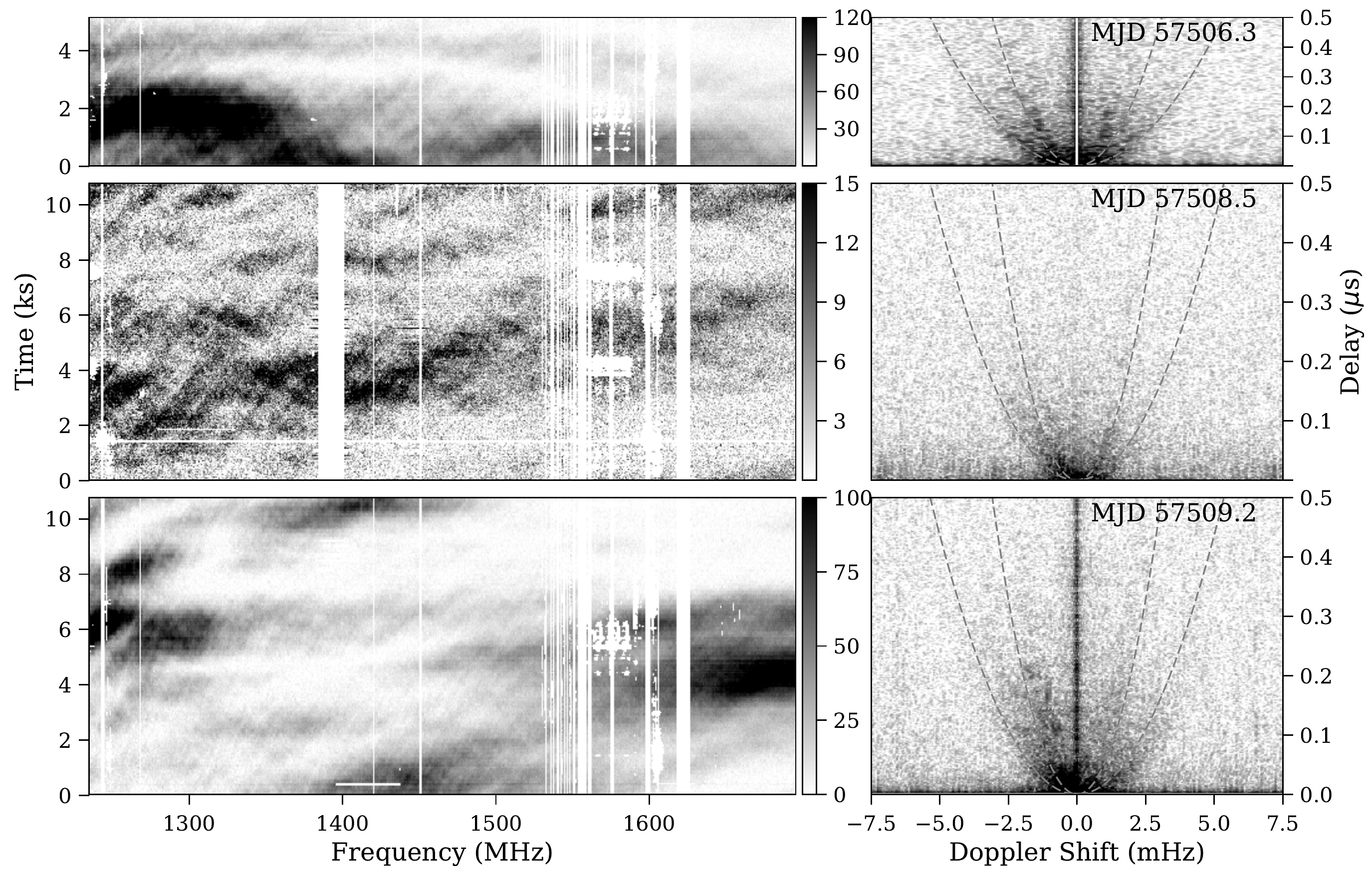}
  \caption{\label{fig:j1057_hoh_1}Left: A sequence of dynamic spectra
    of PSR~J1057$-$5226 taken with the H-OH receiver on three
    consecutive days.  Right: The corresponding secondary spectra.
    Note that the power along the axes of 0\,mHz and 0\,$\mu$s is
    largely due to spectral leakage (see \S\ref{sec:data}), and we
    have pre-whitened/post-darkened the spectra for MJDs 57506.3 and
    57509.2.  The parabolae are drawn with $a=0.02$ (outer) and
$a=0.06$\,s$^{3}$ (inner).} \end{figure*}

\begin{figure*}
\includegraphics[angle=0,width=0.95\textwidth]{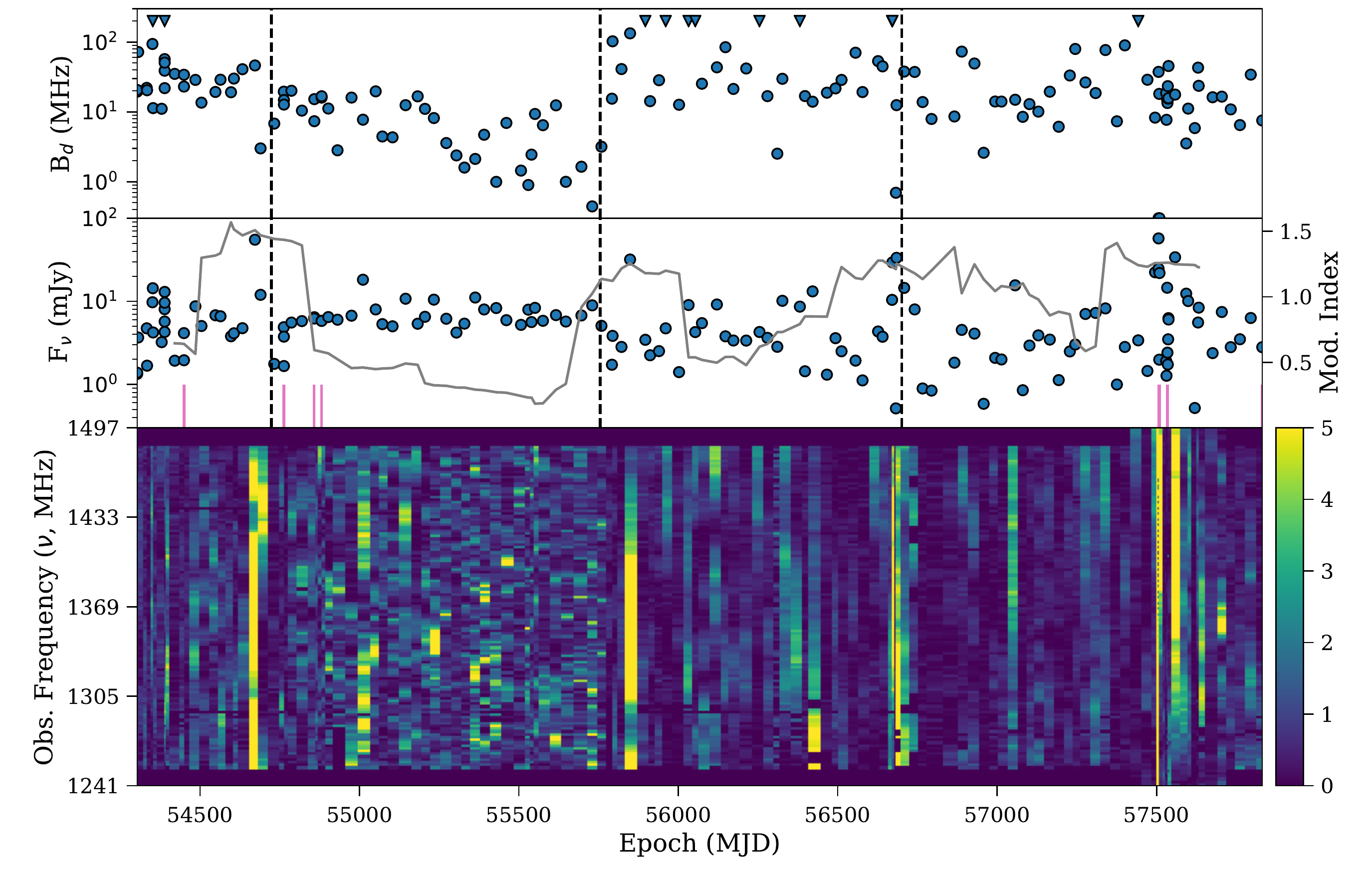}
\caption{
\label{fig:j1057_ds}The dynamic spectra, integrated flux density, and
measured scintillation bandwidth for PSR~J1057$-$5226 over the full
observing span.  Top panel: Scintillation bandwidth measurements,
obtained by fitting an exponential model to the secondary spectra (see
$\S$\ref{sec:j1057_ese}).  Middle panel: Band-averaged flux densities at
each epoch and the corresponding modulation index.  Both here in in the bottom panel, in order to put all
observations on the same footing, we have averaged in time over the
full observation or over the first three minutes, whichever is
shorter.  The horizontal red lines show the epochs of the long
observations whose secondary spectra we show in Figures
\ref{fig:j1057_arcs_2}, \ref{fig:j1057_hoh_1}, \ref{fig:j1057_hoh_2},
and \ref{fig:j1057_new}.  The
gray trace shows the modulation index obtained from observations taken
within six months of the indicated epoch, and consequently no
measurements appear for epochs within six months of
the ends of the observing span.  Bottom panel: the pulsar
spectra from our snapshot observations.   The spectra have been
normalized to the best-fit mean power law model
$F_{\nu}=7.04\,(\nu/1.4\,\mathrm{GHz})^{-1.09}$\,mJy.  To better
display the large dynamic range, the scale saturates at 5, but
outliers range up to about 20.
}
\end{figure*}

\section{Observations and Data Reduction}
\label{sec:data}

We observed PSRs~J1057$-$5226 and J1740$-$3015 as part of a continuing
program to time about 150 young (characteristic ages
$10^{5\pm 0.5}$\,yr), energetic
($\dot{E}>10^{34}$\,erg\,s$^{-1}$) pulsars with the Parkes radio
telescope to provide rotational ephemerides for $\gamma$-ray analysis
with the \textit{Fermi} Large Area Telescope.  The program has been
extremely successful both in its original purpose \citep[][nearly 40
$\gamma$-ray pulsars discovered using Parkes data]{Abdo13_2pc} and
through the study of the wealth of pulse profiles and timing data
\citep[e.g.][]{Rookyard17,Brook16,Antonopoulou15,Karastergiou11}.  The
data set, now spanning more than eight years, allows detection of
phenomena with time-scales of several years, e.g. DM variations
\citep{Petroff13} and the recently discovered quasi-periodic
modulations in the pulse times of arrival of eight members of the
sample \citep{Kerr16}.  Many of the pulsars in the sample have large
DMs and are (very) strongly scattered by the ISM.  As pulsars become
more strongly scattered, their diffractive time-scale (the period over
which the observed intensity in a narrow band becomes decorrelated)
becomes shorter while their refractive time-scale (over which the
line-of-sight crosses the scattering disk) becomes longer, due to the
increased size of the scattering disk, reaching months to years for
typical lines of sight in the Galactic plane.  These time-scales can
be measured by studying the correlations in pulsar light curves
\citep[e.g.][]{Kaspi92}, prompting us to produce flux-density
calibrated light curves from our timing observations as follows.

The pulsar sample was observed with a roughly monthly cadence with
short, two or three-minute snapshot observations using primarily the
centre pixel of the 20\,cm multi-beam receiver.  For about six months
in 2007 and in 2016, we used another 20\,cm (``H-OH'') receiver.  We
recorded 256\,MHz of bandwidth centred on 1369\,MHz using a succession
of digital polyphase filterbanks systems (PDFB1-PDFB4) and folded the
resulting 1024 spectral channels into pulse profiles with 1024 phase
bins.   (With the H-OH receiver in 2016, we recorded 2048 channels
over 512\,MHz of bandwidth at 1465\,MHz.)  To measure the differential
gains between the signal paths of the two voltage probes, once an hour
we observe a pulsed noise signal injected into the signal path prior
to the first-stage low-noise amplifiers.  The noise signal also
provides a reference brightness for each observation.  To correct for
the absolute gain of the system, we make use of observations of the
radio galaxy 3C\,218 (Hydra A); by using on- and off-source pointings,
we can measure the apparent brightness of the noise diode as a
function of radio frequency.  Finally, to measure the flux density, we
cross-correlate the observed profile with a standard template.  For
the total flux density at each epoch, we have integrated over the
observing band and observation duration, while for the dynamic spectra
we retain the native resolution of the filterbanks.  All data
reduction makes use of the \textsc{psrchive} \citep{Hotan04} software
package.

Dynamic spectra are particularly sensitive to radio-frequency
interference (RFI) as both quasi-stationary narrowband and impulsive
broadband signals can create spurious enhancements of frequency and
time intervals.  To mitigate RFI and aliasing, we filter out the edges
of the band (5\%) and channels exceeding a median-smoothed bandpass,
and we identify sub-integrations affected by impulsive RFI by eye and
remove them.  These missing channels and sub-integrations introduce
additional windowing to the secondary spectra discussed below, but the
percentage of filtered data is typically small.  When constructing
secondary spectra, we linearly interpolate over missing data to
minimize windowing artifacts.  Large features in dynamic spectra such
as bright scintles manifest as low-frequency power in secondary
spectra and, due to windowing from finite observation length, can leak
to higher power.  We partially mitigate this in affected observations
by first-order differencing (pre-whitening) and subsequent correction
of the output spectrum (post-darkening); see \citet{Coles11} for more
details.

\section{PSR~J1057$-$5226}
\label{sec:j1057}

PSR~J1057$-$5226 (B1055$-$52) is a bright pulsar discovered in the
seminal, early pulsar search survey of the southern sky by
\citet{Vaughan72}.  Its high S/N profile has furnished several
insights into the pulsar mechanism \citep{Weltevrede12} and it is one
of the brightest $\gamma$-ray pulsars in the sky.  Indeed, the
constraints from its radio polarimetry and $\gamma$-ray light curve
together provide some of the first evidence for ``backwards''
$\gamma$-ray beams \citep{Craig15}.  PSR~J1057$-$5226 has also been
detected at optical and UV wavelengths using the \textit{Hubble} Space
Telescope, allowing sensitive astrometry and refined spectral distance
estimates as we review below.

\subsection{Distance and Velocity}

The interpretation of scintillation measurements depends critically on
the distance and velocity of a pulsar, but unfortunately the distance
to PSR~J1057$-$5226 is not well constrained.  According to the
electron distribution model of \citet{Taylor93}, the modest DM
(30.1\,pc\,cm$^{-3}$)
places it at a distance of 1.53\,kpc, but more modern models place it
at 0.72\,kpc \citep[NE2001][]{Cordes02} and 94\,pc
\citep[YMW16][]{Yao17}, an order of magnitude discrepancy.
Constraints from other observations place the pulsar in the
lower half of this range.  \citet{Mignani10}, by modelling of the
optical, UV, and X-ray spectrum, estimate a distance of 400\,pc.  The
high $\gamma$-ray efficiency, 25.6\% at 400\,pc \citep{Abdo13_2pc},
also suggests a true distance at the lower end of this range, and we
therefore adopt a fiducial distance of 400\,pc.

\citet{Mignani10} measure a proper motion of $42\pm5$\,mas yr$^{-1}$
by comparing the \textit{Hubble} position with an archival radio
timing position.  At our fiducial distance, this leads to a transverse
velocity of 80\,km\,s$^{-1}$, somewhat low but consistent with typical
two-dimensional speeds \citep[median $\sim$180\,km\,s$^{-1}$,][]{Hobbs05}.
However, due to timing noise and the relatively short span of data
used in producing the archival radio position, we sought to verify the
proper motion using a single long span of pulsar timing data.

Archival data from Parkes observations are available with varying
cadence from 1992 and include a range of instruments and receivers.
The raw data are currently unavailable, so we used a partially-reduced
version of the data that were folded at the correct pulsar period and
reduced to, typically, 4 or 8 spectral channels.  Ultimately, we
assembled 2306 pusle times of arrival (TOAs) spanning from July 1992
to January 2016.  The noise properties of the TOAs vary strongly over
the data span.  Radiometer noise generally decreases over time as more
sensitive, wider bandwidth systems became available.  Early data
sometimes suffer pulse profile distortions due to single-bit
digitization, and nearly all lack polarization calibration.  To
account for this noise, we flagged each unique combination of receiver
and backend system and modified the TOA errors with an overall scale
and an additional error added in quadrature (``EFAC'' and ``EQUAD'').
We also allow each system to have a free phase adjustment (``JUMP'')
relative to the fiducial system (PDFB4).  Finally, we fit the timing
model and these white noise parameters, together with a power law
representation of red timing noise, using \textsc{TempoNest}
\citep{Lentati14}.  At an epoch of MJD 54531, we find a position
\begin{equation*}
  \alpha=10^{\mathrm{h}}57^{\mathrm{m}}58\fs959(4),\,\delta=-52\degr26\arcmin56\farcs42(4)
\end{equation*}
and a proper motion
\begin{equation*}
  \mu_{\alpha}\cos\delta = 44\pm\,5\,\mathrm{mas\,yr}^{-1},\,\mu_{\delta}
  = -6\pm\,5\,\mathrm{mas\,yr}^{-1},
\end{equation*}
with a total proper motion of $44\pm5$\,mas\,yr$^{-1}$, verifying
the result of \citet{Mignani10}.  Unfortunately, the precision of the
parallax measurement is only 60\,mas, so we can only place a lower
limit on the distance of about 10\,pc.

\subsection{Scintillation Properties}

Refraction contributes most strongly at the transition from weak to
strong scattering \citep[e.g.][]{Coles87}, which occurs when the
transverse coherence scale in the scattering medium becomes smaller
than the Fresnel radius, or equivalently when the decorrelation
bandwidth becomes smaller than the observing frequency $\nu_0$.
Though depending on frequency and the wavenumber spectrum, refraction
contributes of order 0.1 \citep{Rickett90} to the modulation index
($m\equiv\sigma_{F}/F$, with $F$ the observed flux density) while
diffractive scintillation has $m=1$.  As we show below, we observed
$m=1.2$ for J1057$-$5226 outside of the ESE, indicating a strong
$m=0.7$ for the refractive component.  Thus, even though the observed
decorrelation bandwidth for our 20\,cm observations is $<\nu_0/20$,
formally placing it well into strong scattering, refraction plays an
outsize r\^{o}le, perhaps indicating the presence of an inner scale in
the turbulence spectrum \citep{Coles87}.  Such strong variations may
also reflect the presence of discrete refracting structures.  The
observational consequence of this large variance is that we
occasionally see flux densities almost 20 times greater (or lower)
than the mean even though our bandwidth encompasses multiple
diffractive scintles.

Identifying the ``baseline'' diffractive properties of
PSR~J1057$-$5226 is challenging.  \citet{Johnston98} report values of
36.7\,MHz (correlation half-width at half-maximum) and 308\,s
(correlation half-width at 1/$e$) at 1388\,MHz.  However, long
observations almost always reveal fringing in the dynamic spectra
characteristic of interference between rays scattered from large
angles (see below), and these fringes obscure the underlying
large-scale scintles from the scattering core.  We have identified a
few long observations that appear relatively free from such
interference (e.g. the top panel of Figure \ref{fig:j1057_arcs_2}) and
estimate a slightly larger decorrelation bandwidth of 50\,MHz and
decorrelation time-scale of 1000\,s, which we adopt for a fiducial
model.

We can check the consistency of this model by noting that, if the
pulsar velocity is dominant, the decorrelation time-scale is simply
the time it takes a coherent patch on the scattering screen to
translate past the line-of-sight.  If the phase variations are due to
a Kolmogorov spectrum of fluctuations, then the coherence length
$s_d^2=\zeta(1-\zeta)/2 (B_d/\nu_0) r_f^2$, with $\zeta$ the normalized
distance from the pulsar to the screen, $r_f=\sqrt{\lambda D/2\pi}$
the Fresnel distance, and $B_d$ the diffractive bandwidth.  We report
these parameters in Table \ref{tab:j1057_fiducial} and, if we place
the screen halfway to the pulsar, we find the coherence length to be
$4.4\times10^9$\,cm.  If the effective velocity along the screen is
dominated by the pulsar velocity, i.e. is 40\,km\,s$^{-1}$, then the
predicted decorrelation time is 1100\,s, in good agreement with our
model.  The corresponding refraction timescale is 66 hours, somewhat
longer but in qualitative agreement with the day-to-day variations we
observe (e.g. Figure \ref{fig:j1057_hoh_1}).

\begin{figure*}
  \includegraphics[angle=0,width=0.95\textwidth]{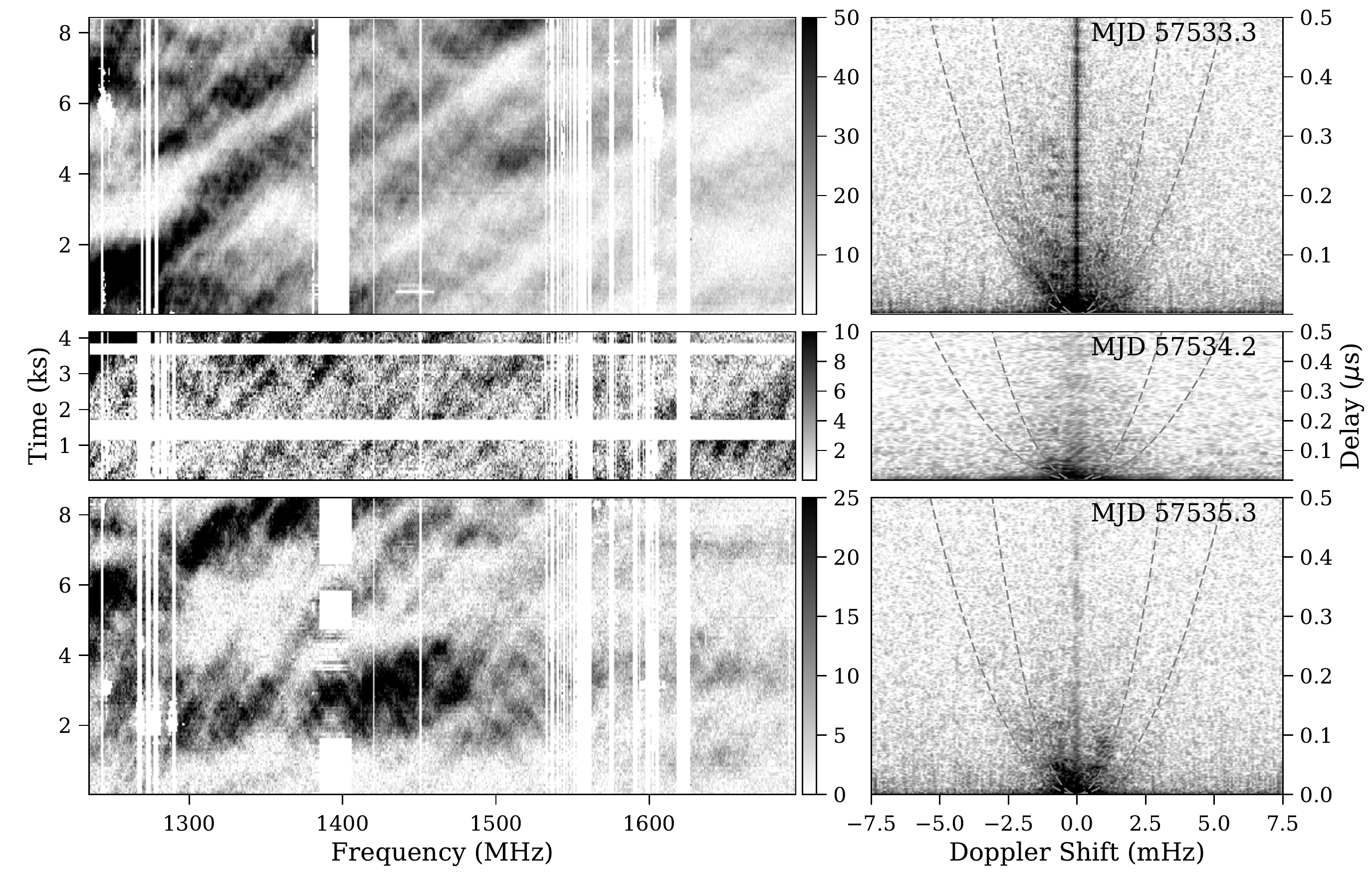}
  \caption{\label{fig:j1057_hoh_2}A sequence of PSR~J1057$-$5226
    dynamic and secondary spectra taken with the H-OH receiver on
    three consecutive days.  We pre-whitened/post-darkened the
    secondary spectrum from MJD 57533.3.  Parabolae are drawn with
    $a=0.02$ (outer) and $a=0.06$\,s$^{3}$ (inner).}
\end{figure*}

\begin{figure*}
  \includegraphics[angle=0,width=0.95\textwidth]{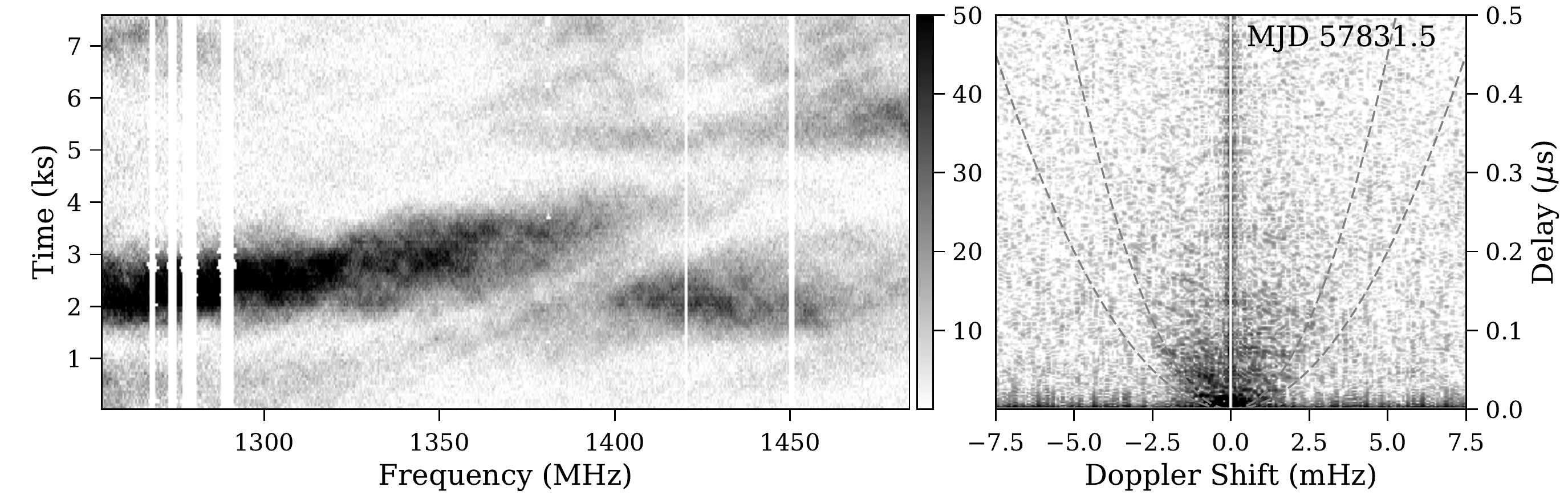}
  \caption{\label{fig:j1057_new}PSR~J1057$-$5226 dynamic and secondary
    spectrum (with pre-whitening/post-darkening) made in March, 2017
    with the multi-beam receiver.  Parabolae are drawn with $a=0.02$
(outer) and $a=0.06$\,s$^{3}$ (inner).} \end{figure*}

\subsection{Extreme Scattering Event}
\label{sec:j1057_ese}

We have examined the flux density and diffractive properties of
J1057$-$5226 over our data set and show our results in Figure
\ref{fig:j1057_ds}.  To measure the decorrelation bandwidth ($B_d$),
we computed the secondary spectrum for each observation (the Fourier
transform of the dynamic spectrum) and then fitted the models
used by \citet{Coles15}.  We have also measured the flux density for
each observation.  The typical observation length (120 to 180\,s) is
much shorter than the decorrelation time-scale (1000\,s), so they give
snapshots whose variance is dominated by the diffractive
scintillation structure.  This is borne out by the running modulation
index, shown in the middle panel of the figure, which reaches values
up to $\sim$1.2.  In order to put the longer observations on the same footing,
we simply truncate those observations to 180\,s when measuring the
flux density.%  We display these results, along with 256-channel
%spectra obtained from the snapshot observations, in Figure
%\ref{fig:j1057_ds}.

There is a substantial drop in both the decorrelation bandwidth and
the variance in the snapshot flux density beginning around MJD 54725
and concluding around MJD 55755, shown as the first two vertical
dashed lines in Figure \ref{fig:j1057_ds}.  Both are hallmarks of an
increase in the scattering strength, and thus the ESE is due to
\emph{additional} turbulent plasma entering the line-of-sight.  (The
flux variance decreases as there are more scintles in the band for
stronger scattering, so the snapshot observations approach the mean
flux.)  There are strong, wideband flux density enhancements preceding
and following the ESE.  These may be coincidental, or they may be
refractive enhancements (lensed emission) associated with the boundary
of the scattering structure.  The scattering structure itself cannot
lens strongly, as there is no measurable decrease in the mean
observed flux.

There is a clear gradient in the diffractive bandwidth over the
duration of the ESE, with the scattering strongest in the trailing
edge.  The sharp delineation in diffractive bandwidth makes the event
duration a relatively precise 1000$\pm$50\,d.  However, its linear
size depends on where it lies along the line of sight.  If we take it
as halfway to pulsar ($\zeta=1/2$), and again assign the dominant
motion to the pulsar velocity ($V_{\mathrm{eff}}=40$\,km\,s$^{-1}$),
the transverse size is 23\,($V_{\mathrm{eff}}/40$\,km\,s$^{-1}$)\,au.
This is substantially larger than structures reported for other ESEs,
even if it is much closer with a typical local relative velocity of
10\,km\,s$^{-1}$.

%(Can anything be said about the
%nature of the scattering since we have this trend?  NB it's similar to
%the trend for the ESE of J1017$-$7156.  More generally, can anything
%be said about the column density needed to produce this?)

%TODO -- if we assume a similar DM profile to the MSP ESEs, then we could calculate the gradient at the sharp edges and see if the lensing angle matches up with the refractive features we see.

We have marked an additional event at MJD 56700, a series of very
bright observations bookending a near non-detection.  These may also
represent a small plasma lens crossing the line of sight, with
characteristic caustic enhancements preceding and following a
defocused region \citep[e.g.][]{Clegg98}.  The decorrelation bandwidth
appears to decrease slightly following this event.

There are additional less ``extreme'' but similar events.  Around MJD
56300 the drop in modulation index is accompanied by a single
observation with a markedly decreased scintillation bandwidth,
reminiscent of (but in the opposite sense to) the ``pinhole'' in DM
reported by \citet{Coles15} for PSR~J1713$+$0747.

%transitions = [54724,55755,56701]

\subsection{Large-angle Scattering}

Although fringing in pulsar dynamic spectra had been appreciated for
some time as a realization of multiple imaging \citep{Rickett90}, it
was the discovery of dramatic parabolic arcs in secondary spectra
\citep{Stinebring01} that crystallized the understanding that these
features are the natural hallmarks of a medium capable of supporting
wide-angle scattering \citep{Walker04,Cordes06} and that they are a
sensitive probe of the angular scattering spectrum.  Briefly, in the
case of interference between a dominant small-angle scattering core
and a halo of light scattered at larger angles, the scattered waves
will be Doppler shifted ($f_t$) by the screen velocity an amount
proportional to the scattering angle and will suffer a delay
($f_{\nu}$) proportional to its square.  The scattering halo sites
then map onto a parabola
\begin{equation}
\label{eq:parab1}
 f_{\nu}=a f_t^2
 \end{equation}(see equation 11 of \citet{Cordes06}, NB in the second
 equality the signs of the exponents of the velocity and units are
 inverted), with
 \begin{equation}
\label{eq:parab2}
 a=0.116\times4\,\zeta(1-\zeta)\,D\,\nu^{-2}
 V_{\mathrm{eff}}^{-2}\,s^3,
 \end{equation}
with $D$ the pulsar
distance in kpc, $\nu$ the observing frequency in GHz, and
$V_{\mathrm{eff}}$
the projected screen velocity scaled by 100\,km\,s$^{-1}$.  Therefore, in
addition to the snapshot spectra discussed above, we have analyzed
long archival observations, taken primarily in 2007 and 2008, as well
as recent data obtained after the
discovery of the ESE.  These long observations provide the necessary
resolution to identify scattering features in secondary spectra.

A very long observation in 2007 (Figure \ref{fig:j1057_arcs_2}, top
panel) gives a reasonable representation of the ``baseline''
scattering state of PSR~J1057$-$5226.  Large structures in frequency
and time, with a strong refractive enhancement in the first part of
the observation, yield a dense knot of scattering near the origin in
the secondary spectrum; the dimensions of this knot yield the
canonical diffractive scintillation bandwidth and time-scale.  The
next long archival observations occur in 2008, after the onset of the
ESE.  The first such observation, the second from the top, shows
obvious fringing, and the secondary spectrum reveals these fringes as
power well removed from the origin, with negative Doppler shifts of a few
mHz and delays of up 0.1\,$\mu$s, while the subsequent two
observations show features with delays up to 0.25\,$\mu$s.  For our
fiducial model with $D=0.4$, $\zeta=1/2$, and
$V_{\mathrm{eff}}=40$\,km\,s$^{-1}$, $a=0.02$\,s$^3$, shown as dashed
arc drawn in Figure \ref{fig:j1057_arcs_2} and in good agreement with
the observed curvature.  This agreement reinforces our fiducial
distance and velocity model and again suggests the ESE is $>$20\,au.
in transverse dimension.

%on... [WHAT; I'm surprised one can't get a larger value of $a$ without
%resulting to a more distant screen.  Look into this.]
%  Note that $a$
%  depends very strongly on distance since our pulsar velocity estimate
%  relies on proper motion.  Thus, the approximate agreement again
%  suggest our distance solution is close to correct.

We have observed similar, if less pronounced, fringing in recent
observations taken as a dedicated followup to the discovery of the
ESE.  In a series of observations made on subsequent days in late
April/early May 2016 (Figure \ref{fig:j1057_hoh_1}), there is evidence
both of strong refractive enhancement and of scattering structure
apparently associated with an even greater value of $a=0.06$\,s$^3$.
The scintillation arc evolves rapidly on a 1-day time-scale, with the
power along the arc at positive Doppler shift on the first day fading
over the two subsequent days.  The time required for this material to
translate far enough to escape the scattering region is about 50\,d,
so this variability must reflect the scattering process itself.
Indeed, \citet{Coles10} found such variation in arc asymmetry on
refractive time-scales in simulations of Kolmogorov turbulence.

Another series of consecutive long observations at the end of May 2016
shows a similar evolution (Figure \ref{fig:j1057_hoh_2}) .  The first
observation reveals scattering at large delay (0.3\,$\mu$s) and small
negative Doppler shift (-1\,mHz), and there is extremely strong
low-frequency fringing.  Two days later, that structure has largely
vanished, while substantial power at small scattering angles, along
the inner parabola, has appeared.  A final long observation made in
March of 2017 (Figure \ref{fig:j1057_new}) reveals a symmetric
distribution concentrated at the origin, suggesting absence of
large-angle interference.

If the larger value of $a$ is a distinct feature, and if the effective
velocity remains dominated by the pulsar velocity, then the screen
distances are related by $a_1/a_2 =
\zeta_1\,(1-\zeta_2)/\zeta_2\,(1-\zeta_1)$.  If $\zeta_1=1/2$ during
the ESE is correct, then $\zeta_2=3/4$.  That is, the ``baseline''
screen is located 1/4 of the way from the earth to the pulsar, and
halfway between the ESE screen and the earth.  Since there is also
still power filling the area between the two parabolae (particularly
during the strong fringing on MJD 57533), this suggests scattering
occurs over an extended region from 100--200\,pc.

Finally, under the assumption that the scattering is dominated by
bright core/large angle interference, the secondary spectrum can be
directly mapped to a brightness distribution.  For our fiducial model,
a parabola following $a=0.02$\,s$^3$ yields an approximately circular
brightness distribution, while smaller (larger) values are elongated
along the (perpendicular to the) effective velocity vector.  Thus, the
brightness distribution is circular during the ESE (save for the
bright large angle knot) but somewhat elongated perpendicular to the
velocity in subsequent epochs.  In fact, change of image shape may
also be sufficient to explain the different apparent values of $a$, as
asymmetry in the image strongly affects the position of scattered
power relative to the predicted parabola \citep[e.g.][]{Cordes06}.

\begin{table}
\caption{\label{tab:j1740_fiducial}Fiducial scattering model values for
PSR~J1740$-$3015 for observing frequency 1369\,MHz and screen at
$\zeta=1/2$.}
\begin{tabular}{lr}
\hline
Distance ($D$, kpc) \dotfill & 2.0 \\ 
Dispersion Measure (DM, $e^{-1}$\,pc\,cm$^{-3}$) \dotfill & 152 \\ 
Transverse velocity (km s$^{-1}$) \dotfill & 150 \\
Fresnel radius (cm) \dotfill & $1.5\times10^{11}$ \\
Decorrelation bandwidth ($B_d$, MHz) \dotfill & 1 \\
Decorrelation time-scale (s) \dotfill & -- \\
Diffractive scale ($s_d$ cm) \dotfill & $1.4\times10^9$ \\
Refractive scale (cm) \dotfill & $1.6\times10^{13}$ \\
Diffractive time-scale (s) \dotfill & 190 \\
Refractive time-scale (d) \dotfill & 24 \\
\end{tabular}
\end{table}

\begin{figure*}
\includegraphics[angle=0,width=0.95\textwidth]{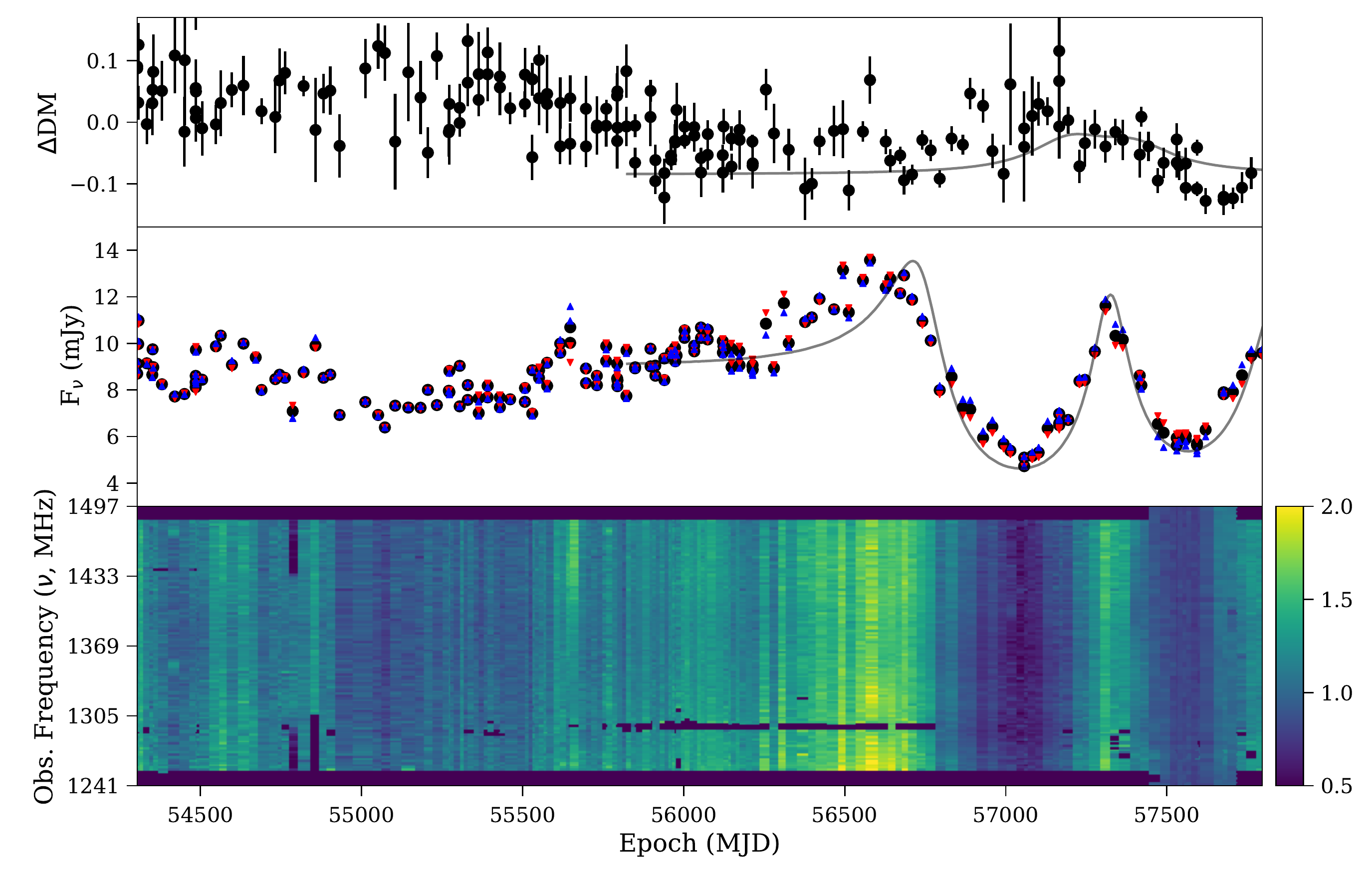}
\caption{\label{fig:j1740_ds}The DM variation, light curve, and
  dynamic spectra for PSR~J1740$-$3015 over the full observing span.
  Top: The DM as estimated from in-band pulsar timing.  The gray trace
  shows the predicted DM from the simple lens model described in
  \S\ref{sec:j1740_ese}.
Middle: The band-averaged flux density (black circles) and the flux
density for the top (blue upward-pointing triangles) and bottom (red
downward-pointing triangles) halves of the band after scaling for the
mean spectral index.  The gray trace shows the flux variations
predicted by the simple lens model.
Bottom: the dynamic spectrum, averaged over time, for each observing
epoch.  The spectra have been normalized to the best-fit mean power
law model $F_{\nu}=7.58\,(\nu/1.4\,\mathrm{GHz})^{-1.05}$\,mJy.}
\end{figure*}

\section{PSR~J1740$-$3015}
\label{sec:j1740}

PSR~J1740$-$3015 (B1737$-$30), discovered by \citet{Clifton86} at a DM
of 152\,pc\,cm$^{-3}$, is a radio-bright but otherwise unremarkable pulsar, save for
its relatively frequent glitches \citep{Clifton86}.  Its precise
distance remains unknown.  \citet{Johnston01b} were able to detect
H\textsc{i} absorption at 15\,km\,s$^{-1}$, but the 3\,kpc arm is seen only in emission, leading those authors to conclude
only that the pulsar is nearer than $5.5\pm0.6$\,kpc.
\citet{Verbiest12} more recently invoked Malmquist bias to suggest
that the distance is only 0.4$^{+1.7}_{-0.3}$\,kpc.  Both
the NE2001 and the YMW16 DM/distance models predict intermediate
distances of 2.7 and 2.9\,kpc, respectively.  Due to the lack of any
obvious strongly-ionized regions along the line-of-sight and the small
decorrelation bandwidth (see below), we favour these intermediate
distances and adopt a fiducial pulsar distance of 2.0\,kpc, marginally
consistent with the large uncertainties of \citet{Verbiest12}.
Unfortunately, there are no constraints on the pulsar proper motion,
and we adopt a typical speed of 150\,km\,s$^{-1}$.

We note that PSR~J1740$-$3015 lies at an ecliptic latitude of
$-$6.9\degr, and consequently in December passes very close to the
sun, with some observations having a solar angle $<10\degr$.  Such
observations (about ten in our set) suffer a degraded system
temperature and additional bandpass ripple from standing waves between
the telescope focus and antenna surface.  However, we find no evidence
that the post-calibration spectra are out of trend and we do not
exclude these data.

\subsection{Scintillation Properties}

Measurements of the correlation bandwidth, as well as inspection of
Figure \ref{fig:j1740_ds}, indicate the scintillation bandwidth is
$\sim$1\,MHz, making the measurement via correlation of channels of
marginal reliability.  We also evaluated the scintillation bandwidth
by measuring excess variance in the dynamic spectra.  In brief, the
total variance within a sub-integration should be equal to the
radiometer contribution plus a $\chi^2_{2n}$, with $n$ the number of
scintles within the band.  Because intrinsic variation in pulse
amplitude and shape (``jitter'') is typically correlated over
bandwidths much larger than the observing bandwidth \citep{Shannon14},
its noise contribution is negligible.  The typical excess variance,
corresponding to a value of $n=220$ in a 256\,MHz band, confirms a
$\sim$1\,MHz bandwidth.  The baseline model values with our fiducial
distance and velocity are given in Table \ref{tab:j1740_fiducial}.

After removing the low-frequency trend from the measured flux
densities (middle panel of Figure \ref{fig:j1740_ds}), the
band-integrated modulation index is about 0.07.  The expected variance
from DISS is 5--10\%, indicating residual contributions from
refraction at a 24-day time-scale are minimal.  However, there is
clearly variation on longer time-scales.  Due to the low ecliptic
latitude, the orbital motion of the earth may be substantial relative
to the scattering medium, leading to an annual signal.  However, when
we computed a power spectrum of the long-term light curve (by
interpolating and pre-whitening/post-darkening to correct for spectral
leakage) we found no clear quasi-periodic signal.

\subsection{Extreme Scattering Event}
\label{sec:j1740_ese}

Above the level of refractive variation, there is a clear increase in
received flux density beginning at MJD~56250, though the leading edge may
extend earlier in time.  Relative to a baseline flux density of
8\,mJy, the ESE peaks around 13\,mJy (163\%), followed by a dip to
5\,mJy (61\%).  A second more moderate (11.5\,mJy) peak and trough
(5.8\,mJy) follow.  In a typical diverging lens geometry, at least one
more peak is expected.  During the ESE, there appears to be a
low-frequency ripple in the intensity across the observing band
(bottom panel), with greater intensity towards the edges and a clear
dip in in the band center; this variation is most clearly visible near
the first intensity peak at MJD~56600.  We have also plotted the flux
density for the top and bottom halves of the observing band, scaled to
the mean spectral index, as triangles in the middle panel.  These
light curves show that the ESE is frequency-independent over our
observing band.%  As we discuss
%below, these properties are not consistent with the simple
%($\propto\lambda^2$) dependence expected from a refractive lens.  

This is the longest ESE observed to date.  \citet{Maitia03} report a
three-year ESE towards the millisecond pulsar J1643$-$1224, but this
is likely a special case of a persistent annual modulation related to
the earth's orbital motion \citep{Keith13}.  The duration of this ESE
is already at least 1550\,d, and if the expected second peak is
similar in shape to the first, the total length will be around
2000\,d.

\begin{figure}
\includegraphics[angle=0,width=0.50\textwidth]{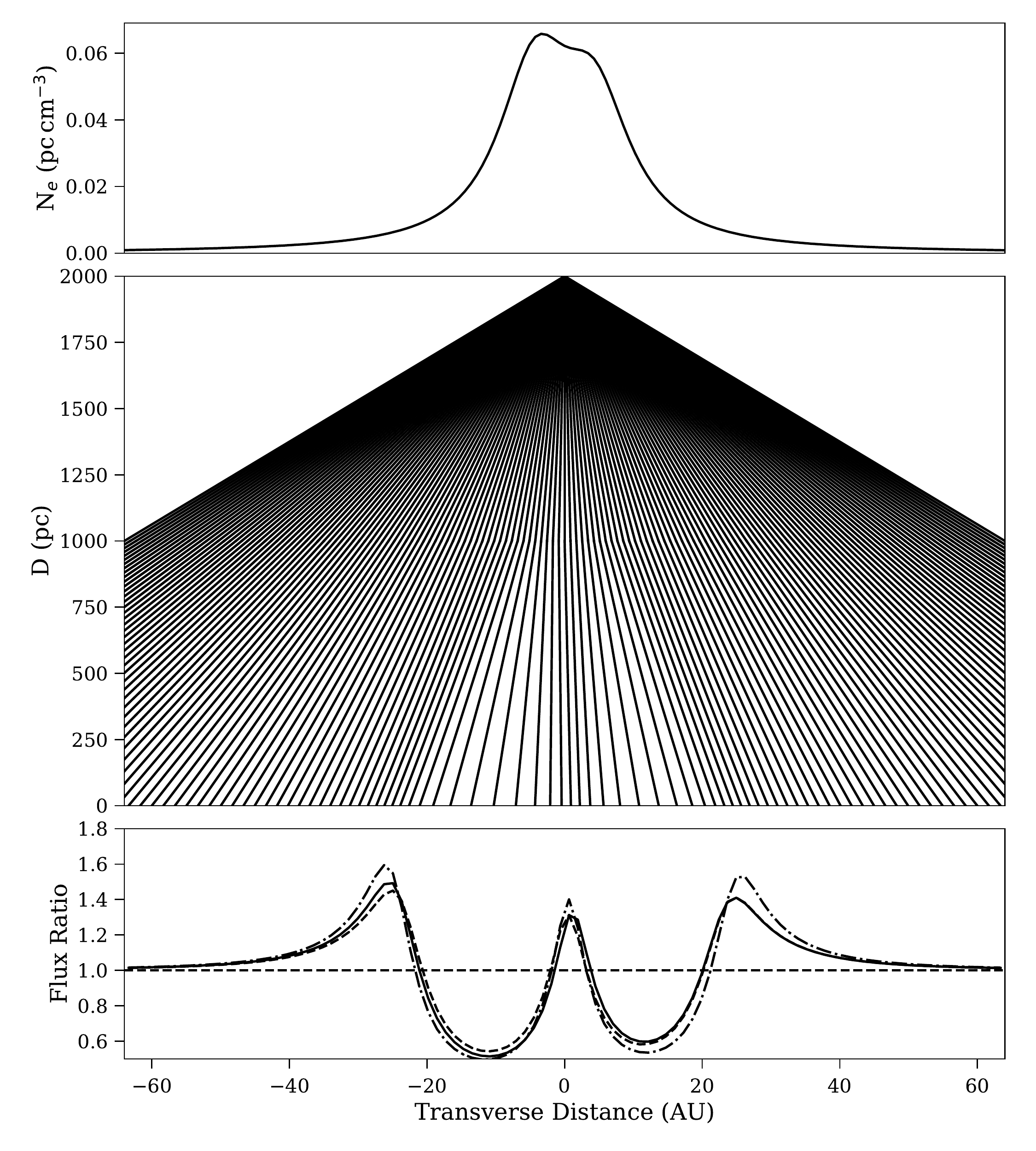}
\caption{\label{fig:j1740_lens}A diverging plasma lens model for the
J1740$-$3015 ESE.  Top: the predicted electron column density in typical
DM units.  Middle: The evolution of the field in the geometric
approximation.  Bottom: The received intensity.}
\end{figure}

Following \citet{Clegg98}, we have modelled the flux density
modulation as a diverging plasma lens crossing the line-of-sight.  The
two dips require two modes in the distribution, and we compute the
received flux density using simple ray tracing, taking into account
the finite source distance.  The resulting model, at our fiducial
screen distance of 1000\,pc, appears in Figure \ref{fig:j1740_lens}.
In this model, the two Lorentzian components with HWHM $w$ have
separation 0.68\,$w$ and the trailing component column density N$_e$
amplitude is 80\% of the first.  We note that Gaussian components
provide a poorer fit as modulation between the peaks is too deep.
With the effective screen velocity of 75\,km s$^{-1}$, the predicted
transverse size of $\sim$80\,au is in good qualitative agreement with
the data.  The finite distance of the source means the inferred lens
size is smaller than one in the plane wave approximation.

This model is not unique.  Even setting aside more complicated (e.g.
2-d) lens shapes, there is degeneracy between the lens
distance, electron column density (N$_e$), and size \citep{Clegg98},
and here a two-parameter family of lenses
provides a good description of the data.  The assumed transverse
velocity fixes the lens size at a given distance to be
$w=6.4$\,au\,(D$_l$/1\,kpc)($V_0$/75\,km\,s$^{-1}$), and the peak/trough
amplitude then requires the normalization column density to follow
$N_e=0.050$\,\,pc\,cm$^{-3}$(D/1\,kpc)($V_0$/75\,km\,s$^{-1}$)$^2$.  Here,
$V_0$ represents the normalization of the transverse screen velocity,
scaled in such a way that the observed ESE length is correct at the
other model values.  This scaling has two important consequences.

First, we obtain equally good results over a broad range of lens
distances.  However, small lens distances can be ruled out because the
image scale becomes too small to account for the long ESE with
reasonable relative velocities.  More distant lenses require greater
electron column depths, and for lenses further than about 1\,kpc the
required column would exceed our observational constraints on DM
variation.  In the top
panel of Figure \ref{fig:j1740_ds}, we show the DM as computed by
fitting with \textsc{Tempo2} a $\nu^{-2}$ dispersion model to 8 TOAs
extracted across the observing band along with the DM column for our
fiducial model.

The second consequence is that the electron density is quite robustly
determined for a lens at any valid distance.  For our model, the FWHM
of the density profile is about $3w$.  If we take this as
characteristic of the longitudinal dimension, the peak electron
density $n_e\approx500\,\mathrm{cm}^{-3}(V_0/75\,\mathrm{km\,s}^{-1})$
depends only on the velocity scale.  This electron density is somewhat
lower than the values estimated for shorter ESEs, in keeping with the
larger inferred size.

Finally, we note that the 1-d plasma lens model predicts chromaticity
as the refraction angle $\propto\lambda^2$.  We show the difference in
flux densities for the two halves of our observing band as the
non-solid lines in Figure \ref{fig:j1740_lens}.  Although the
predicted differences are relatively modest over our bandwidth, they
are detectable, suggesting a more complicated lens is necessary to
fully account for the data.%  We note that 2-d lenses can produce
%arbitrary dynamic spectra.

\section{Discussion}
\label{sec:discussion}

\begin{figure}
\includegraphics[angle=0,width=0.45\textwidth]{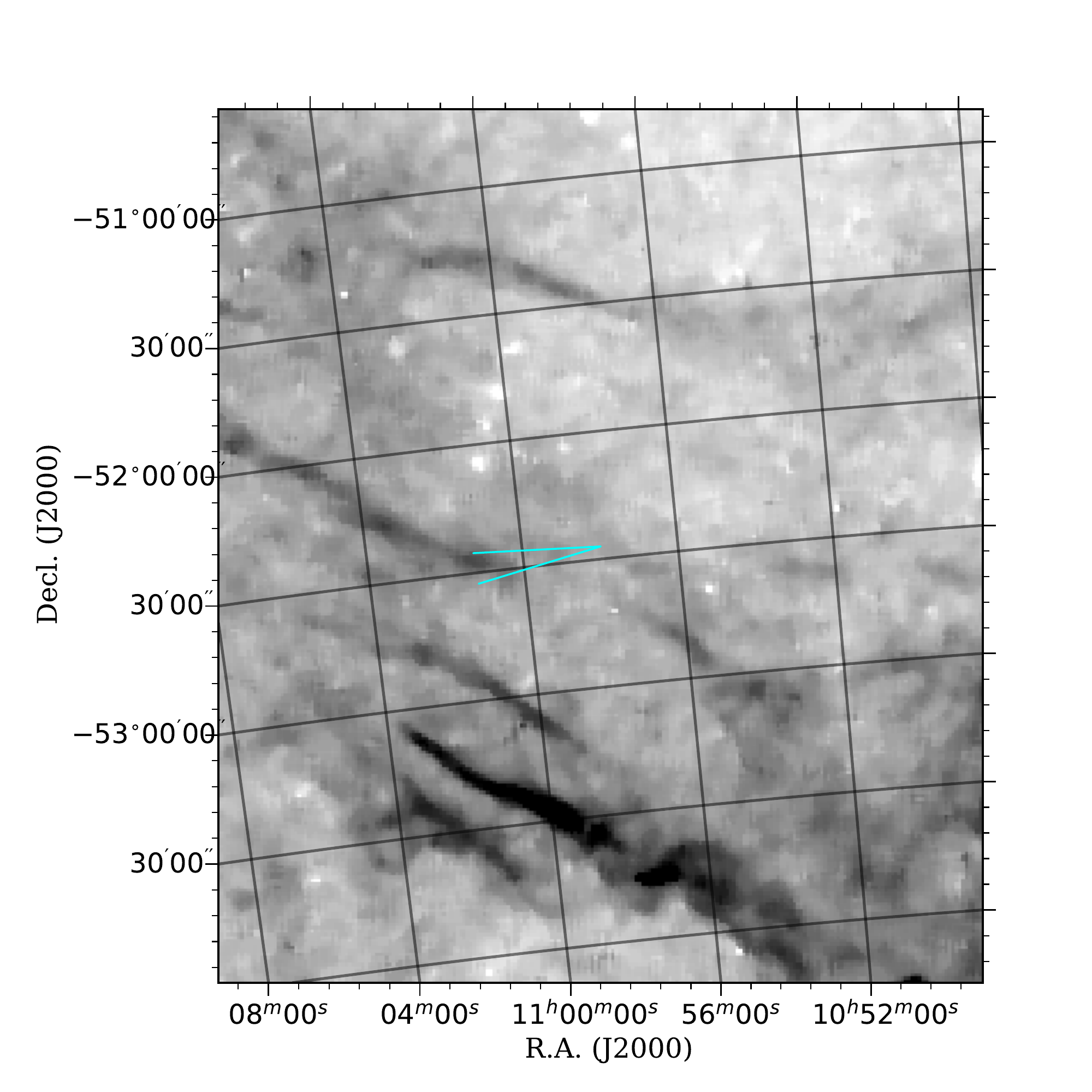}
\caption{\label{fig:j1057_halpha}An H$\alpha$ image from SHASSA
  \citep{Gaustad01}
showing the pulsar location and the cone containing its velocity
vector from the measurement of \citet{Mignani10}.}
\end{figure}

\citet{Romani87} proposed two mechanisms to explain the observations
of \citet{Fiedler87}: sheets of ionized gas in old supernova remnants
whose pressures could confine sheets with $n_e\sim50$\,cm$^{-3}$,
sufficient to provide the observed enhancement if their aspect ratios
are $\sim$100; and very high pressure ($n_e\sim10^4$\,cm$^{-3}$)
magnetically-confined plasma spun off by stars.  Both ideas have
re-appeared recently in more generalized form.  The likely ubiquity of
current sheets in the ISM \citep{Braithwithe15} has been invoked both
to explain ESEs \citep{Pen12} and more generally pulsar scintillation
\citep{Pen14}.  The newly-discovered strong correlation between the
velocity of some intra-day variables (IDVs) and hot stars likewise
lends support for a stellar origin of scattering structure
\citep{Walker17}.  A unifying feature of all of these origins is
substantial elongation either along or perpendicular to the line of
sight, and the ESEs reported here provide further evidence for such
asymmetry.

As we have described above, the ESEs towards PSR~J1057$-$5226 and
J1740$-$3015 have long durations and transverse sizes of
23\,($V_{\mathrm{eff}}/40$\,km\,s$^{-1}$)\,au and
80\,($V_{\mathrm{eff}}/75$\,km\,s$^{-1}$)\,au, respectively.  For
PSR~J1057$-$5226, the good agreement between our fiducial model and
the curvature of the secondary spectrum parabola suggests the
transverse size truly is at least 20\,au.  For J1740$-$3015, our
fiducial model is consistent with a simple plasma lens model,
indicating a size $>$20\,au, but a wider range of values is equally
good.  Nonetheless, it is probably larger than 5\,au, as a closer
(smaller) lens would likely show pronounced annual effects from the
earth's orbit.  These large sizes---and the lack of shorter/smaller
ESEs in our sample---suggest the ESEs are indeed the result of
inhomogeneities rather than properties of a self-similar turbulent
medium.

These large transverse sizes have interesting implications for the
models above.  If the scatterer is sheet-like, then only inclinations
close to edge-on provide the strong gradients needed to provide the
refraction in PSR~J1740$-$3015 and the sharp edges for
PSR~J1057$-$5226.  Such alignment occurs by chance of order 1\% of the
time, consistent with the detection rates reported here (2 events/7
ESE-years per 150 pulsars/1200 source-years) and those of
\citet{Fiedler94}).  Further fine tuning such that the elongation is
along the effective velocity yields detection rates lower than
observed, suggesting that the scattering structures are roughly
symmetric in the plane of the sky and overall geometry is an elongated
cylinder.  If the scatterer is a corrugated plasma sheet, then
the corrugations are much larger than the sheet width.

There are several candidates for such extended structures along the
line of sight to PSR~J1057$-$5226.  It is near the boundary of the
classical Loop 1 bubble, and more convincingly along the intersection
of two shells proposed by \citet{Wolleben07} to explain the north
pulsar spur, with distances comparable to our estimate for the screen.
We also note that \citet{Yao17} place the Loop 1 shell at 195\,pc,
precisely at our fiducial screen distance.  Inspection of an H$\alpha$
image (Figure \ref{fig:j1057_halpha}) reveals filamentary structure on
large scales, lending credence to the idea that the region may harbor
such post-shock electron density enhancement as those proposed by
\citet{Romani88} and \citet{Clegg88}.  On the other hand, we note that
the ESE cannot be associated with an underdensity
\citep[e.g.][]{Pen12} since the scintillation strength increases.
 
Intriguingly, several type A Hipparcos stars \citep{Perryman97} lie
both in the foreground of our 400\,pc distance and within $\sim$1\,pc
of the sight line.  HIP53771 (px=16.4(4)\,mas, 1/px=60\,pc,
$\rho$=0.7\,pc, A3III/IV, with $\rho$ the distance from the sight
line) is the strongest candidate for providing a scattering radial
plasma structure \citep{Walker17}.  Two other A stars, HIP53395
(px=8.5(3)\,mas, 1/px=120\,pc, $\rho$=0.9\,pc, A9V) and HIP53836
(px=7.3(4), 1/px=140\,pc, $\rho$=1.2\,pc, A0V) are also candidates.
In all cases, the surrounding ionized volumes are substantially
smaller than their Hill spheres, so it is unclear if they are close
enough.  Future precise astrometry from GAIA, as well as a monitoring
campaign to search for annual signatures in diffraction properties,
can further test the hot star hypothesis. 

Unfortunately, due to the larger and more uncertain distance to
PSR~J1740$-$3015, less can be said about specific scatterers.  There
are not many hot stars near earth and near the sight line, with only
two Hipparcos candidates.  HIP86744 (px=8.1(3)\,max, 1/px=120\,pc,
$\rho$=1.3\,pc, A2V) and HD316162 (px=0.8(3), 1/px=1200\,pc, 2.8\,pc,
A0) could plausibly provide structure along the sight line.  A third
Hipparcos source, HIP 86512 (A0 RR Lyr variable), has a revised GAIA
distance of $\sim$1\,kpc \citep{gaiadr1} which places it too far from
the sight line.  Generally, we look forward to discovering more
candidates with further GAIA data releases.% and HIP 85340 (b Oph) is
%a nearby (40\,pc) A5 giant but about 3\,pc from the sight line, again
%likely too far.  Although GAIA DR2 may yield more candidates, hot
%stars seem an unlikely explanation for this ESE.

Although these ESEs clearly illuminate some properties of the ISM,
much more could be learned, and potentially some models discarded,
with additional observational data.  In particular, both candidates
would benefit from wideband monitoring.  E.g., the ultra-wideband
receiver currently being commissioned at Parkes (0.7--4.0\,GHz) would
expand the band to half the current observing frequency and thus will
substantially increase the number of scintles observable in an
observation of PSR~J1057$-$5226.  This may better characterize the
refractive properties of future ESEs as well as yield better
constraints on the parabolic arcs in the secondary spectrum.
Conversely, the decorrelation bandwidth for PSR~J1740$-$3015 will be
resolvable above 3\,GHz with even modest channelization, allowing for
searches of annual modulation in diffraction which might allow
association with a nearby star.  In both cases, modelling the
frequency evolution of the diffractive properties will allow better
understanding of the turbulence spectrum dominating the scattering.

Likewise, a VLBI parallalax would greatly expand our understanding of
these sources.  Knowing the pulsar distance and proper motion would
break the degeneracy in the lens model for J1740$-$3015, allowing
determination of the screen location.  Determing the distance of
J1057$-$5226 would determine whether local bubble boundaries are
the most likely scatterers and would give a direct measurement of the
screen velocity from the parabolic arcs.  Since both pulsars are
bright, the measurement is possible with existing observatories.

Finally, we close with a note about the dramatic flux variations
observed towards PSR~J1057$-$5226 in its ``quiescent'' state, outside
of the ESE.  We have observed flux densities in excess (deficit) of
the mean by factors of 20.  Although the pulsar is not far off the
Galactic plane (6.7\degr), the scattering is likely dominated by local
material and thus many extragalactic sight lines could experience such
``super'' refractive scintillation if the sources are sufficiently
compact.  Monitoring campaigns of nearby
pulsars over a range of frequencies and many refractive time-scales
will reveal just how often such modulation can happen.  Such
substantial magnifications may play a r\^{o}le in the detectability of
fast radio bursts \citep{Macquart15} and better determine the relation between FRBs and
the local ISM, if indeed any exists \citep{Fiedler94}.

\section*{Acknowledgements}

We thank Phil Edwards for his deft telescope scheduling.  The Parkes
radio telescope is part of the Australia Telescope, which is funded by
the Commonwealth Government for operation as a National Facility
managed by CSIRO.  This research has made use of NASA's Astrophysics
Data System, for which we are grateful.  Work at NRL is supported by
NASA.

\input{ms.bbl}
%\bibliographystyle{mn2e}
%\bibliography{../sr_whitepaper_2013/sr}

%\appendix

\label{lastpage}

\end{document}